\documentclass[%
 aip,
 amsmath,amssymb,
 reprint,%
]{revtex4-1}

\usepackage{graphicx}
\usepackage{dcolumn}
\usepackage{bm}

\usepackage[utf8]{inputenc}
\usepackage[T1]{fontenc}
\usepackage{mathptmx}
\usepackage{etoolbox}

\makeatletter
\def\@email#1#2{%
 \endgroup
 \patchcmd{\titleblock@produce}
  {\frontmatter@RRAPformat}
  {\frontmatter@RRAPformat{\produce@RRAP{*#1\href{mailto:#2}{#2}}}\frontmatter@RRAPformat}
  {}{}
}%
\makeatother
\begin{document}

\preprint{AIP/123-QED}

\title[]{Fluid viscoelasticity suppresses chaotic convection and mixing due to electrokinetic instability}
\author{C. Sasmal}%
 \email{csasmal@iitrpr.ac.in}
\affiliation{ 
Soft Matter Engineering and Microfluidics Lab, Department of Chemical Engineering, Indian Institute of Technology Ropar, Punjab, India-140001.
}

\date{\today}

\begin{abstract}
When two fluids of different electrical conductivities are transported side by side in a microfluidic device under the influence of an electric field, an electrokinetic instability (EKI) is often generated after some critical values of the applied electric field strength and conductivity ratio. Many prior experimental and numerical studies show that this phenomenon results in a chaotic flow field inside a microdevice, thereby facilitating the mixing of two fluids if they are Newtonian in behaviour. However, the present numerical study shows that this chaotic convection arising due to the electrokinetic instability can be suppressed if the fluids are viscoelastic instead of Newtonian ones. In particular, we observe that as the Weissenberg number (ratio of the elastic to that of the viscous forces) gradually increases and the polymer viscosity ratio (ratio of the solvent viscosity to that of the zero-shear rate viscosity of the polymeric solution) gradually decreases, the chaotic fluctuation inside a T microfluidic junction decreases within the present range of conditions encompassed in this study. We demonstrate that this suppression of the chaotic motion occurs due to the formation of a strand of high elastic stresses at the interface of the two fluids. We further show that this suppression of the chaotic fluctuation (particularly the span-wise one) inhibits the mixing of two viscoelastic fluids. Therefore, one needs to be cautious when the EKI phenomenon is planned to use to mix such viscoelastic fluids. Our observations are in line with that seen in limited experimental studies conducted for these kinds of viscoelastic fluids.     
\end{abstract}

\maketitle

\section{\label{sec:level1}Introduction}
Over the past three decades or so, numerous microdevices based on the electrokinetic (EK) transport mechanism have been developed, resulting in an extensive research interest from a wide range of research communities on this topic. These small-scale devices are often used for various purposes, such as sample pretreatment, mixing, separation, etc. They are integrated as a primary component of the so-called micro total analysis systems ($\mu$TAS) used for many chemical and biochemical analyses and detection~\cite{kovarik2012micro,west2008micro,arora2010latest}. The traditional pressure-driven microdevices such as syringe pumps are hardly preferred in these $\mu$TAS. This is because they involve moving mechanical parts, which are often challenging to manage in such small-scale systems. In contrast, EK-based microdevices require no mechanical forces as they rely on applying an electric field. This also results in the design and fabrication of these EK-based microdevices being simpler than the pressure-driven ones. However, many complex flow phenomena often arise in these EK-based microdevices due to the handling of samples of large heterogeneity in their properties and sometimes with the increase in the design complexity of the system. One such example is the electrokinetic instability (EKI) phenomenon that arises due to the presence of an electrical conductivity gradient in the samples under the influence of a high electric field. The gradient in the electrical conductivity in the samples can sometimes occur intentionally, such as in the sample stacking processes, or unintentionally during the handling of multi-dimensional assays. Therefore, a significant body of research efforts has been invested in acquiring an in-depth understanding of this flow phenomenon over the years~\cite{lin2009electrokinetic,oddy2001electrokinetic}. 

The EKI phenomenon is considered one of the electrohydrodynamics (EHD) instability problems wherein the latter subject has been studied extensively since Taylor and Melcher's pioneering work on the leaky Ohmic dielectric model~\cite{melcher1969electrohydrodynamics}. According to this model, originally proposed for macrosystems, the EHD instability is originated due to the charge accumulation at the material junctions. This charge accumulation can directly influence the fluid motion due to its coupling with the momentum equation via the electric body force term. For electrokinetic microscale systems, however, some modifications were needed to this Ohmic model. In particular, two main changes were introduced, namely, the exclusion of the relaxation mechanism of electrolytes as it happens instantaneously and the inclusion of the molecular diffusion process of electrolytes as it becomes essential in these small-scale systems. By performing so, Lin et al.~\cite{lin2004instability} proposed a modified Ohmic model for leaky dielectric fluids and conducted extensive theoretical, numerical, and experimental investigations on this EKI phenomenon in a rectangular cross-sectioned microfluidic channel. They found that the flow became unstable and chaotic as the applied electric field exceeded a critical value. Due to this fluctuating flow field, rapid mixing was observed in this microfluidic device. A good agreement was also seen between their theoretical and experimental analyses. Many other studies also observed the enhancement in the mixing phenomena due to this instability phenomenon. For instance, see the studies by Jin and Hu~\cite{jin2010mixing} for a Y-shaped microchannel, Park et al.~\cite{park2005application} for a T-shaped microdevice with an outlet microchannel having different grooved shapes, Huang et al.~\cite{huang2006application} for a cross-shaped microchannel, etc. On the other hand, Chen et al.~\cite{chen2005convective} performed an extensive investigation, consisting of scaling analysis, simulations, and experiments, on this phenomenon in a microfluidic T-junction device. They found convectively unstable and propagating upstream waves in the flow profile. A linear stability analysis was performed to analyze the unstable mode and the nature of these instabilities. Their scaling analysis and numerical results suggested that two parameters govern these instabilities, namely, the ratio of dynamic to dissipative forces that controls the onset of these instabilities and the ratio of electroosmotic to electroviscous velocities that regulates the relative effect of the convective and absolute instability. Posner et al.~\cite{posner2006convective} conducted a parametric study on this EKI phenomenon in a cross-shaped microchannel with three converging inlets and one outlet using the epifluorescence experimental imaging technique. They found the presence of coherent flow structures of the sinuous and dilational forms when the electrical conductivity gradients were greater than one and less than one, respectively. A detailed analysis of the coherent flow structure in this EKI regime was presented by Dubey et al.~\cite{dubey2017coherent} with the help of the dynamic mode decomposition (DMD) technique.   

Apart from the studies on the influence of the electric field, conductivity ratio, and geometry type, studies were also carried out to investigate the influence of other parameters on the dynamics of this EKI phenomenon. For instance, Luo~\cite{luo2009effect} investigated the effect of the multi-species electrolyte solutions and their injection configurations on the evaluation of this instability phenomenon. Dubey et al.~\cite{dubey2021electrokinetic} performed an extensive study with the help of both simulations and experiments on how these instabilities can be generated and evolved for a system where the directions of both the electrical conductivity gradient and applied electric field are the same. On the other hand, the effect of the channel depth on the onset of this instability was investigated in their parametric three-dimensional numerical simulations of Li et al.~\cite{li2016parametric}. They found an increase in the threshold electric field value with the microchannel depth.

Therefore, from the literature mentioned above, it can be seen that a considerable amount of studies, comprised of scaling arguments, numerical simulations, and experimental analyses, have been conducted to understand the EKI phenomenon. Strikingly, during the literature survey, we noticed that almost all of those studies were carried out by assuming leaky dielectric fluids as simple Newtonian fluids. However, it can be readily acknowledged that many fluids, such as polymer solutions, emulsions, suspensions, foams, etc., are routinely encountered in various micro and nanofluidic systems for their further processing and applications under the influence of an electric field~\cite{pfohl2003trends,nghe2011microfluidics}. Additionally, various biofluids, such as blood, saliva, DNA suspensions, cerebrospinal fluid, etc., are also regularly used in various $\mu$TAS for diagnostics and biochemical analyses. All these fluids do not obey the simple and linear Newtonian constitutive relation. Instead, they exhibit a great extent of complex viscoelastic rheological behavior along with other non-linear behaviors such as shear-thinning, shear-thickening, viscoplasticity, etc~\cite{thurston1972viscoelasticity,davis1971rheological,brust2013rheology,nader2019blood}. Therefore, it is essential to investigate how the rheological behavior of fluid could influence this EKI phenomenon for the better design and performance of a $\mu$TAS that handles these complex fluids frequently. 

To date, only one experimental study, performed by Song et al.~\cite{song2019electrokinetic}, focused on how the viscoelastic behavior of working fluid can influence these instabilities in a T junction microchannel. They observed that the fluid viscoelasticity (induced by the addition of polyethylene oxide (PEO) polymers into a phosphate buffer solution) dramatically alters the critical values of the parameters at which this instability originates in comparison to that seen for a simple Newtonian fluid. Additionally, they also found that the speed and frequency of the convective waves (generated due to these instabilities) are suppressed as compared to those seen in buffer solutions without the polymer additives. They performed their study both with the help of experiments and numerical simulations (neglecting the effect of fluid viscoelasticity) over a wide range of polymer concentrations and electric field strength. However, no explanation was provided on why the difference in the chaotic flow dynamics was seen between the polymer and simple buffer solutions. Therefore, this study aims to fill this gap of understanding by providing an in-depth analysis with the help of numerical simulations by incorporating a proper viscoelastic fluid model into them. In doing so, we plan to use a two-dimensional microfluidic T junction as a model flow system with two converging inlets and one outlet. In particular, in this study, our goal is to address how and why fluid viscoelasticity influences the chaotic flow dynamics and the corresponding mixing phenomena due to this EKI phenomenon.

\section{\label{sec:GovEqs}Problem statement and governing equations}
\begin{figure}
    \centering
    \includegraphics[trim=7cm 5cm 15.5cm 1cm,clip,width=9.5cm]{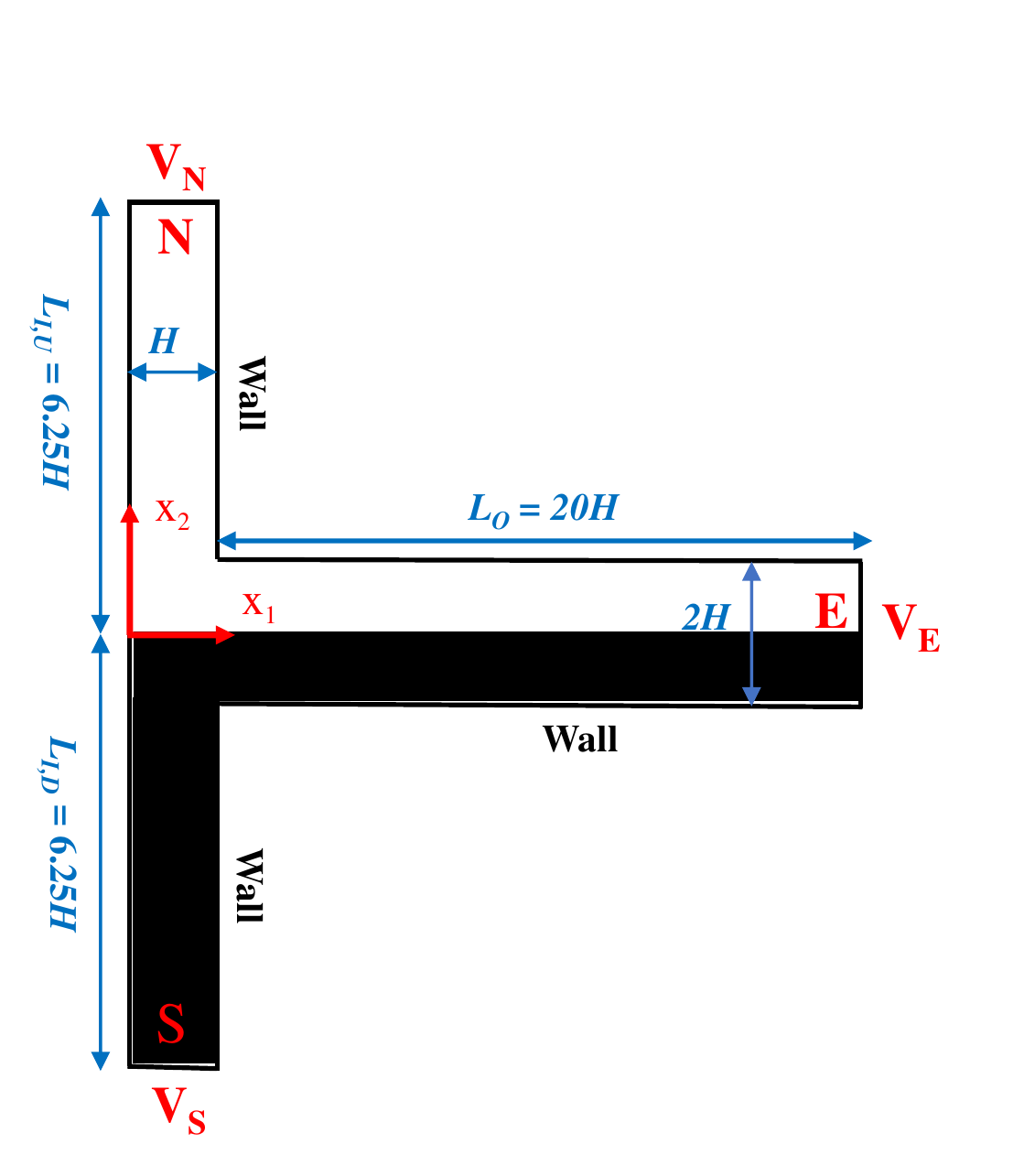}
    \caption{Schematic of the two-dimensional microfluidic T junction used in the present study. Here $2H$ and $H$ are the heights of the outlet and inlet regions of the device, $L_{I,U} = 6.25H$ and $L_{I,D} = 6.25H$ are the lengths of the converging inlets and $L_{O} = 20H$ is the outlet length of the T junction. $V_{N}$, $V_{S}$, and $V_{E}$ are the applied voltages at the north $(N)$, south $(S)$ and east $(E)$ sides of the device, respectively. Here $H = 100 \,\mu m$.}
    \label{fig:1}
\end{figure}
The schematic of the problem considered in this study is shown in Fig.~\ref{fig:1}. It is a two dimensional microfluidic T junction device consisting of two converging inlets and one outlet. While the lengths of the converging inlets are kept as $6.25H$, the outlet length is fixed at $21H$, where $H \, = 100\, \mu m$. The heights of the device in the inlet and outlet sections are $H$ and $2H$, respectively. Leaky dielectric viscoelastic fluids with high and low electrical conductivities enter into the microfluidic T junction through the south $(S)$ and north $(N)$ inlet sections of the device, as schematically shown in Fig.~\ref{fig:1}. 

Depending upon the rheological responses of a viscoelastic fluid in standard homogeneous flows like simple shear or uniaxial extensional flows, one can choose a suitable viscoelastic constitutive relation to account for the fluid viscoelasticity in numerical simulations. In the present study, the Oldroyd-B viscoelastic constitutive equation is used because of the following reasons~\cite{shaqfeh2021oldroyd} i) it is a relatively simple viscoelastic model which depends on a single conformation tensor for predicting the state of stress in a liquid, and associated with only two parameters, namely, polymer concentration and polymer relaxation time ii) it has been derived based on the simplest kinetic theory of polymers in which a polymer molecule is assumed to be a dumbbell with two beads connected by an infinitely stretchable elastic spring iii) it can successfully mimic the rheological behaviour of a constant shear viscosity viscoelastic fluid or the so-called Boger fluid~\cite{james2009boger,bird1987dynamics}. Therefore, the explicit effect of the fluid elasticity on the flow dynamics can be investigated with this constitutive model. The modified Ohmic model derived by Lin et al.~\cite{lin2004instability}, which is the extension of the model proposed by Melcher~\cite{melcher1981continuum}, is used in this investigation to study the electrokinetic instability phenomenon occurring due to the electrical conductivity gradient. This model was developed based on three main assumptions, namely, electroneutrality, negligible diffusive current and symmetric electrolyte. A detailed derivation of this model is already presented by Lin et al.~\cite{lin2004instability} and is also further discussed by Chen et al.~\cite{chen2005convective} and Posner et al.~\cite{posner2006convective}. Therefore, we do not present this again in this study. Under these circumstances, the equations which will govern the present flow dynamics, written in their dimensional forms as follows\newline  
Continuity equation:
\begin{equation}
    \frac{\partial u_{j}^{*}}{\partial x_{j}^{*}} = 0
\end{equation}
Momentum equation:
\begin{equation}
    \rho \left( \frac{\partial u_{i}^{*}}{\partial t^{*}} + u_{j}^{*} \frac{\partial u_{i}^{*}}{\partial x_{j}^{*}} \right) = -\frac{\partial p^{*}}{\partial x_{i}^{*}} + \eta_{s} \frac{\partial}{\partial x_{j}^{*}} \left( \frac{\partial u_{i}^{*}}{\partial x_{j}^{*}} \right) + \frac{\partial \tau_{ij}^{p*}}{\partial x_{j}^{*}} - \rho_{e}^{*}E_{i}^{*}
\end{equation}
Ohmic model equation:
\begin{equation}
 \frac{\partial \sigma}{\partial t^{*}} + u_{j}^{*}\frac{\partial \sigma}{\partial u_{j}^{*}} = D_{\text{eff}}\frac{\partial \sigma}{\partial x_{j}^{*2}}   
\end{equation}
\begin{equation}
    \frac{\partial (\sigma E_{j}^{*})}{\partial x_{j}^{*}} = 0
\end{equation}
\begin{equation}
     \frac{\partial (\epsilon E_{j}^{*})}{\partial x_{j}^{*}} = \rho_{e}^{*}
\end{equation}
Oldroyd-B viscoelastic constitutive equation:
\begin{equation}
         \frac{\partial C_{ij}}{\partial t^{*}} + u_{k}^{*} \frac{\partial C_{ij}}{\partial x_{k}^{*}} = C_{ik} \frac{\partial u_{j}^{*}}{\partial x_{k}^{*}} + C_{kj} \frac{\partial u_{i}^{*}}{\partial x_{k}^{*}} - \frac{1}{\lambda}\left(C_{ij} - \delta_{ij}\right)
\end{equation}
\begin{equation}
    \tau_{ij}^{p*} = C_{ij} -\delta_{ij}
\end{equation}
In the above equations, $\rho$ is the fluid density, $\eta_s$ is the solvent viscosity, $\sigma$ is the electrical conductivity, $\epsilon$ is the electrical permitivity, $E^{*}$ is the electric field, $\rho_{e}$ is the charge density, $x^{*}$ is the position, $u^{*}$ is the velocity vector, $t^{*}$ is the time, $p^{*}$ is the pressure, $\tau^{p*}$ is the polymeric stress tensor, and $\lambda$ is the polymer relaxation time. $D_{\text{eff}}$ is the effective diffusivity which can be calculated for a binary and monovalent fully dissociated electrolyte as $\frac{2D_{+}D_{-}}{D_{+} + D_{-}}$. Here $D_{+}$ and $D_{-}$ are the diffusivities of positive and negative ions, respectively. These can be related to the ionic mobility $\mu_{\pm}$ through the Einstein relation as $D_{\pm} = RT\mu_{\pm}$, where $R$ is the universal gas constant and $T$ is the absolute temperature. $C_{ij} = <R_{i}R_{j}>$ is the conformation tensor where $R_{i}$ is the dimensionless end-to-end vector of a polymer molecule and $\delta_{ij}$ is the Kronecker delta. Note that here $(\,)^{*}$ denotes a dimensional variable.    

The following scaling variables are used to non-dimensionalize the aforementioned governing equations: position with $H$, velocity with $U_{ev}$, time with $\frac{H}{U_{ev}}$, pressure and polymeric stress tensor with $\frac{\eta_{0} U_{ev}}{H}$, electric field with $E_{a}$, and charge density with $\frac{\epsilon E_{a}}{H}$. Here $U_{ev}$ is the electroviscous velocity defined as $\frac{\epsilon E_{a}^{2} H}{\eta_{0}}$ and $E_{a}$ is the apparent applied electric field calculated as $\frac{V_{N} - V_{E}}{L_{O} + L_{I,U}}$, where $\eta_{0}$ is the zero-shear rate viscosity of the polymer solution and $V_{E}$ and $V_{N}$ are the voltages applied at the east and north sides of the device. After performing the non-dimensionalization with these scaling variables, one can obtain the following non-dimensional governing equations
\begin{equation}
    \frac{\partial u_{j}}{\partial x_{j}} = 0
\end{equation}
Momentum equation:
\begin{equation}
    Re \left( \frac{\partial u_{i}}{\partial t} + u_{j} \frac{\partial u_{i}}{\partial x_{j}} \right) = -\frac{\partial p}{\partial x_{i}} + \beta \frac{\partial}{\partial x_{j}} \left( \frac{\partial u_{i}}{\partial x_{j}} \right) + \frac{\partial \tau_{ij}^{p}}{\partial x_{j}} - \rho_{e}E_{i}
\end{equation}
Ohmic model equation:
\begin{equation}
 \frac{\partial \sigma}{\partial t} + u_{j}\frac{\partial \sigma}{\partial u_{j}} = \frac{1}{Ra_{e}}\frac{\partial \sigma}{\partial x_{j}^{2}}   
\end{equation}
\begin{equation}
    \frac{\partial (\sigma E_{j})}{\partial x_{j}} = 0
\end{equation}
\begin{equation}
     \frac{\partial (\epsilon E_{j})}{\partial x_{j}} = \rho_{e}
\end{equation}
Oldroyd-B viscoelastic constitutive equation:
\begin{equation}
         \frac{\partial C_{ij}}{\partial t} + u_{k} \frac{\partial C_{ij}}{\partial x_{k}} = C_{ik} \frac{\partial u_{j}}{\partial x_{k}} + C_{kj} \frac{\partial u_{i}}{\partial x_{k}} - \frac{1}{Wi}\left(C_{ij} - \delta_{ij}\right)
\end{equation}
\begin{equation}
    \tau_{ij}^{p} = \frac{1}{Wi} \left( C_{ij} -\delta_{ij} \right)
\end{equation}
From the above non-dimensional forms of the equations, it can be seen that the present electrokinetic flow phenomena will be governed by the following non-dimensional parameters, namely,
\begin{itemize}
    \item Reynolds number, $Re = \frac{\rho U_{ev} H}{\eta_{0}}$. It is the ratio of the inertial to that of the viscous forces, which is kept low ($\sim$ 2.7) so that the effect of the inertial forces on the flow dynamics can be neglected. 
    \item Electric Rayleigh number, $Ra_{e} = \frac{U_{ev} H}{D_{\text{eff}}}$. It is the ratio of the electroviscous to that of the diffusive velocities. A constant value of $Ra_{e} = 2771.6$ is used in this study. 
    \item Weissenberg number, $Wi = \frac{\lambda U_{ev}}{H}$. It is the ratio of the elastic to that of the viscous forces. A range of values of this dimensionless number is used to examine the influence of the elastic forces on the flow dynamics. Note that for a Newtonian fluid, this number is essentially zero. 
    \item Polymer viscosity ratio, $\beta = \frac{\eta_{s}}{\eta_{0}}$. It is the ratio of the solvent viscosity to that of the zero-shear rate viscosity of the polymeric fluid. For a Newtonian fluid $\beta = 1$, whereas $\beta \rightarrow 0$ represents a polymeric melt. In the present study, a range of values of this parameter is also used to demonstrate the influence of the polymer concentration on the flow dynamics.  
\end{itemize}
Apart from these dimensionless numbers, additionally, the following two dimensionless numbers are also present in this study, namely, conductivity ratio $\gamma = \frac{\sigma_{H}}{\sigma_{L}}$ and voltage ratio $V_{R} = \frac{V_{S}}{V_{N}}$, where $V_{S}$ and $V_{N}$ are the applied voltages at the south and north sides of the microfluidic device as depicted in Fig.~\ref{fig:1}.    

\section{\label{sec:Num}Computational details}
All the governing equations, namely, mass, momentum, Ohmic, and Oldroyd-B viscoelastic constitutive equations, written in the preceding section have been solved using the finite-volume method (FVM) based rheoEFoam solver available in the recently developed RheoTool package~\cite{rheoTool}. This solver has been developed based on the open-source computational fluid dynamics (CFD) code OpenFOAM~\cite{wellerOpenFOAM}. A detailed description of the present solver used in this study is already available elsewhere~\cite{pimenta2018numerical}, and hence only some of the salient features (mainly different discretization techniques) of this solver are recapitulated here. All the advective terms in the governing equations were discretized using the high-resolution CUBISTA (Convergent and Universally Bounded Interpolation Scheme for Treatment of Advection) scheme for its improved iterative convergence properties. All the diffusion terms in the governing equations were discretized using the second-order accurate Gauss linear orthogonal interpolation scheme. All the gradient terms were discretized using the Gauss linear interpolation scheme. The backward time integration scheme was used to discretize the time derivative terms. While the linear systems of the pressure, velocity and electric potential fields were solved using the Geometric-Algebraic Multi-Grid (GAMG) with DIC (Diagonal-based Incomplete Cholesky) preconditioner, the stress, dye concentration and electrical conductivity fields were solved using the Preconditioned Bi-conjugate Gradient Solver (PBiCG) solver with DILU (Diagonal-based Incomplete LU) preconditioner. The pressure-velocity coupling was accomplished using the SIMPLE method, and the log-conformation tensor approach was used to stabilize the numerical solution. Furthermore, the relative tolerance level for the pressure, velocity, stress, dye concentration and electrical conductivity fields was set as 10$^{-10}$. The whole computational domain was discretized using a total of 57000 hexahedral cells. This number was fixed after performing the standard grid independence study at the highest value of the Weissenberg number and lowest value of the polymer viscosity ratio considered in this study where the effect of the fluid viscoelasticity on the flow dynamics would be maximum.  

Finally, the following boundary conditions have been employed in order to facilitate the numerical simulations. For the pressure, a zero $(p = 0)$ condition at all inlet and outlet sides of the device as they are open to atmosphere and a zero gradient $(\frac{\partial p}{\partial n_{i}} = 0)$ at all solid walls are imposed, where $n_{i}$ is the unit outward normal vector. For the electric potential, a constant value $(\psi = C)$ at all inlet and outlet sides of the device depending upon the values of the apparent electric strength $E_{a}$ and the voltage ratio $V_{R}$ and a zero gradient $(\frac{\partial \psi}{\partial n_{i}} = 0)$ at all solid walls are applied . For the velocity, a zero gradient $(\frac{\partial u_{i}}{\partial n_{i}} = 0)$ at all inlet and outlet sides and a slip boundary condition of the form $u_{s,i} = \mu_{0}\frac{\sigma}{\sigma_{0}}^{m}E_{i}$ are implemented~\cite{chen2005convective}. Here $\mu_{0} = -\frac{\epsilon \zeta_{0}}{\eta_{0}}$ (where $\zeta$ is the wall zeta potential) is a reference electroosmotic mobility at a reference electrical conductivity $\sigma_{0}$ and $m$ is an exponent used to account the power-law dependence of the wall zeta potential on the electrical conductivity. A value of $m = -0.3$ is used in this study as suggested by the literature~\cite{chen2005convective}.  

\begin{figure}
    \centering
    \includegraphics[trim=0cm 0cm 0cm 0cm,clip,width=9cm]{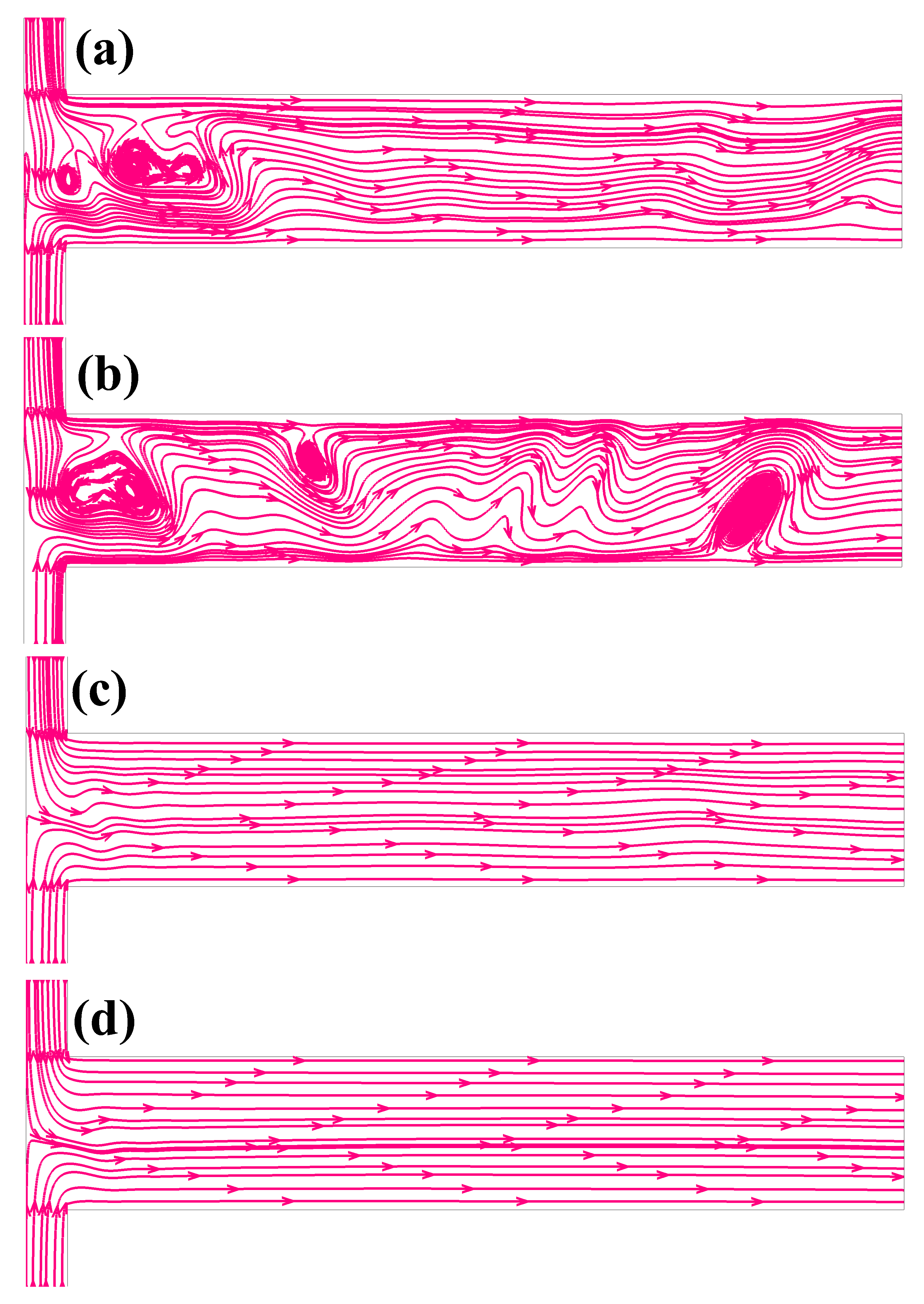}
    \caption{Representative instantaneous streamline patterns inside the microdevice. (a) Newtonian fluid (b-d) viscoelastic fluids with Weissenberg numbers 1, 10 and 15, respectively.}
    \label{fig1}
\end{figure}
\begin{figure}
    \centering
    \includegraphics[trim=0cm 0cm 0cm 0cm,clip,width=7cm]{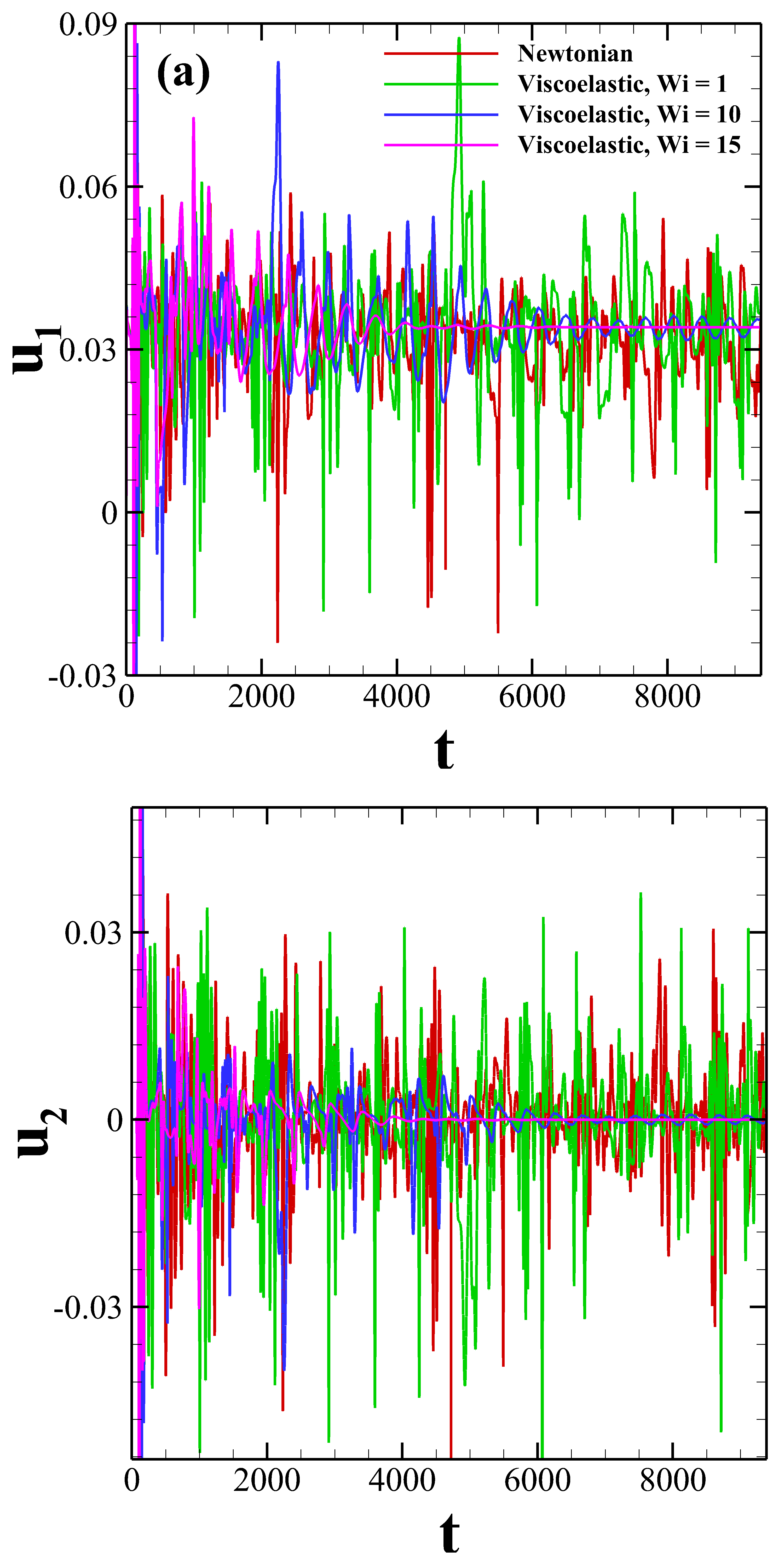}
    \caption{Temporal variation of the (a) stream-wise $(u_{1})$ and (b) span-wise $(u_{2})$ velocities at a probe location placed at $x_{1} = 10H$ and $x_{2} = 0$ both for Newtonian and viscoelastic fluids.}
    \label{fig2}
\end{figure}
\begin{figure*}
    \centering
    \includegraphics[trim=0cm 0cm 0cm 0cm,clip,width=18cm]{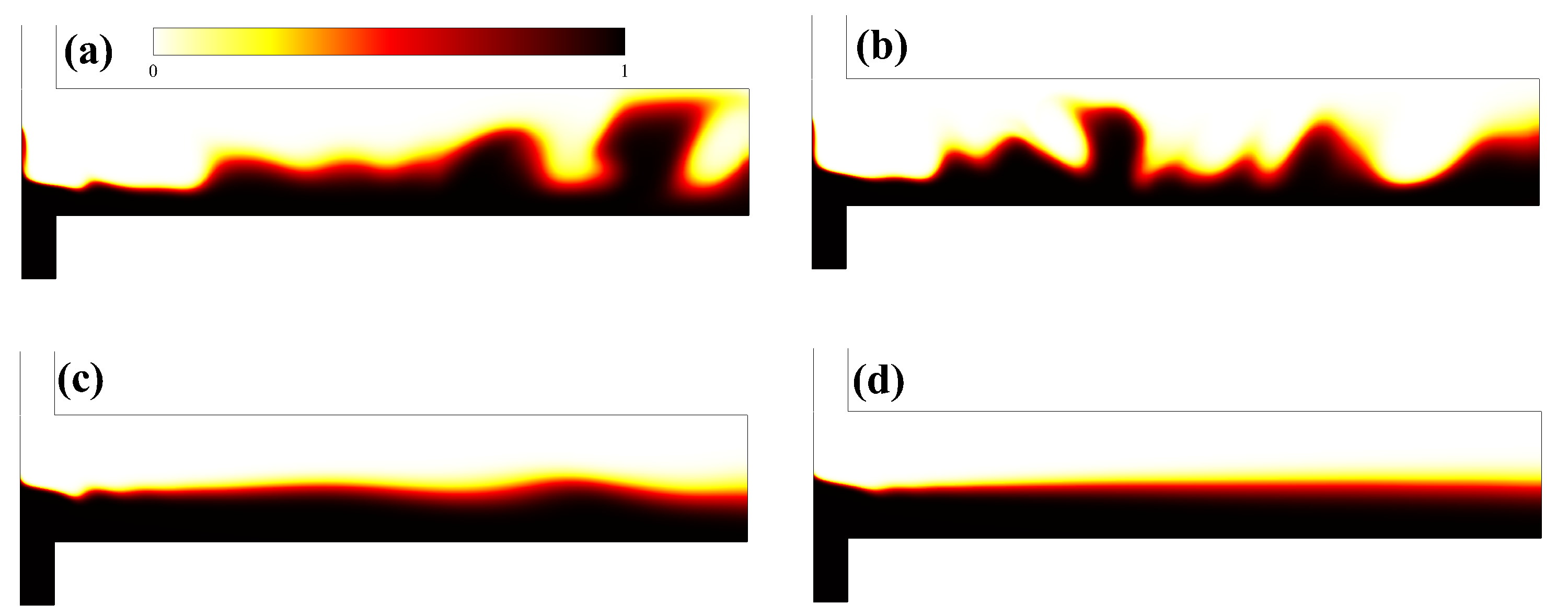}
    \caption{Instantaneous evaluation of the dye concentration profile inside the microdevice. (a) Newtonian fluid (b-d) Viscoelastic fluids with Weissenberg numbers 1, 10 and 15, respectively.}
    \label{fig3}
\end{figure*}
\begin{figure}
    \centering
    \includegraphics[trim=3cm 2cm 1cm 2cm,clip,width=9cm]{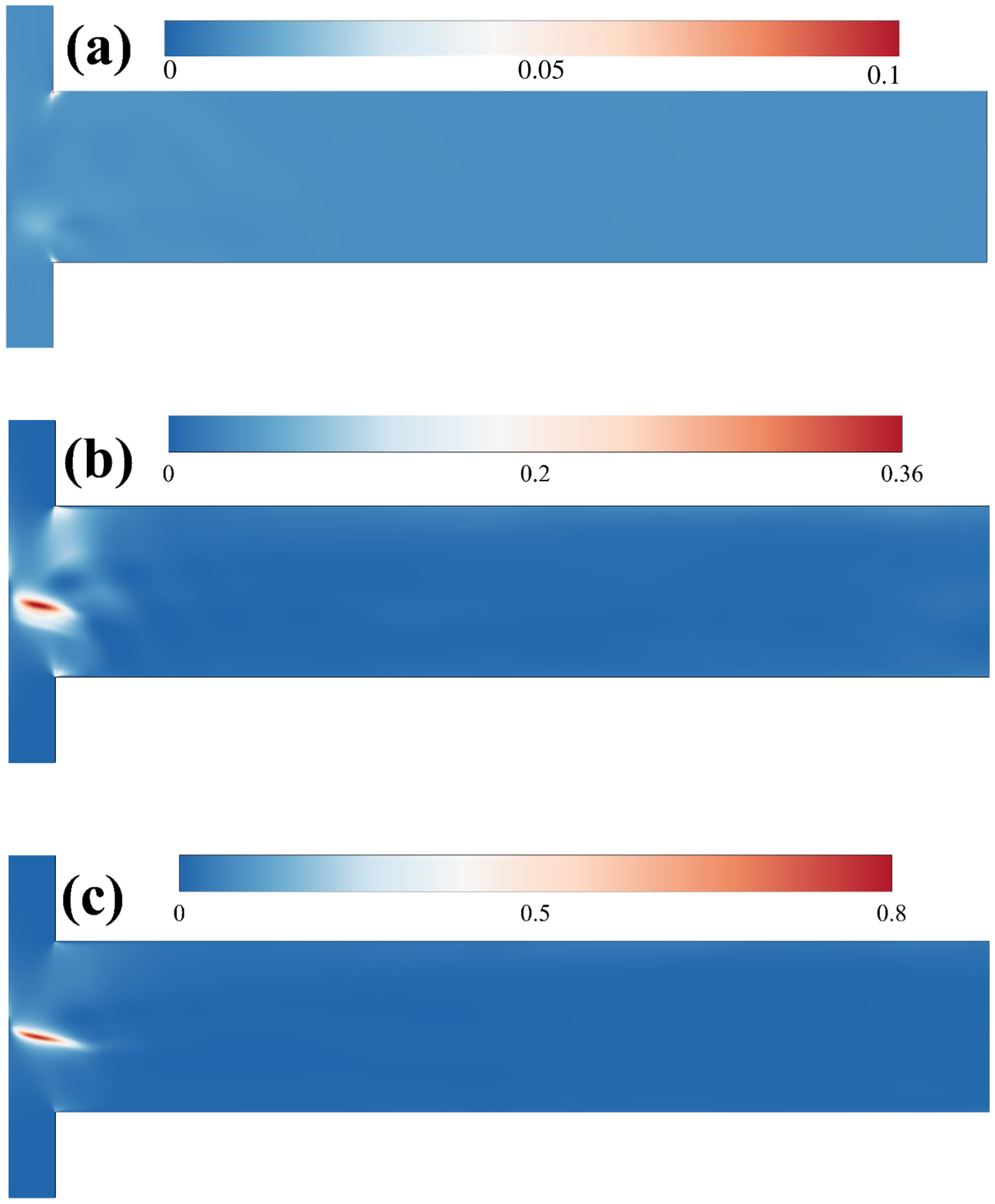}
    \caption{Surface plot of the non-dimensional elastic stress component $\tau_{11}$ for viscoelastic fluid with (a) Wi = 1 (b) Wi = 10 (c) Wi = 15.}
    \label{fig4}
\end{figure}
\begin{figure}
    \centering
    \includegraphics[trim=0cm 0cm 0cm 0cm,clip,width=8cm]{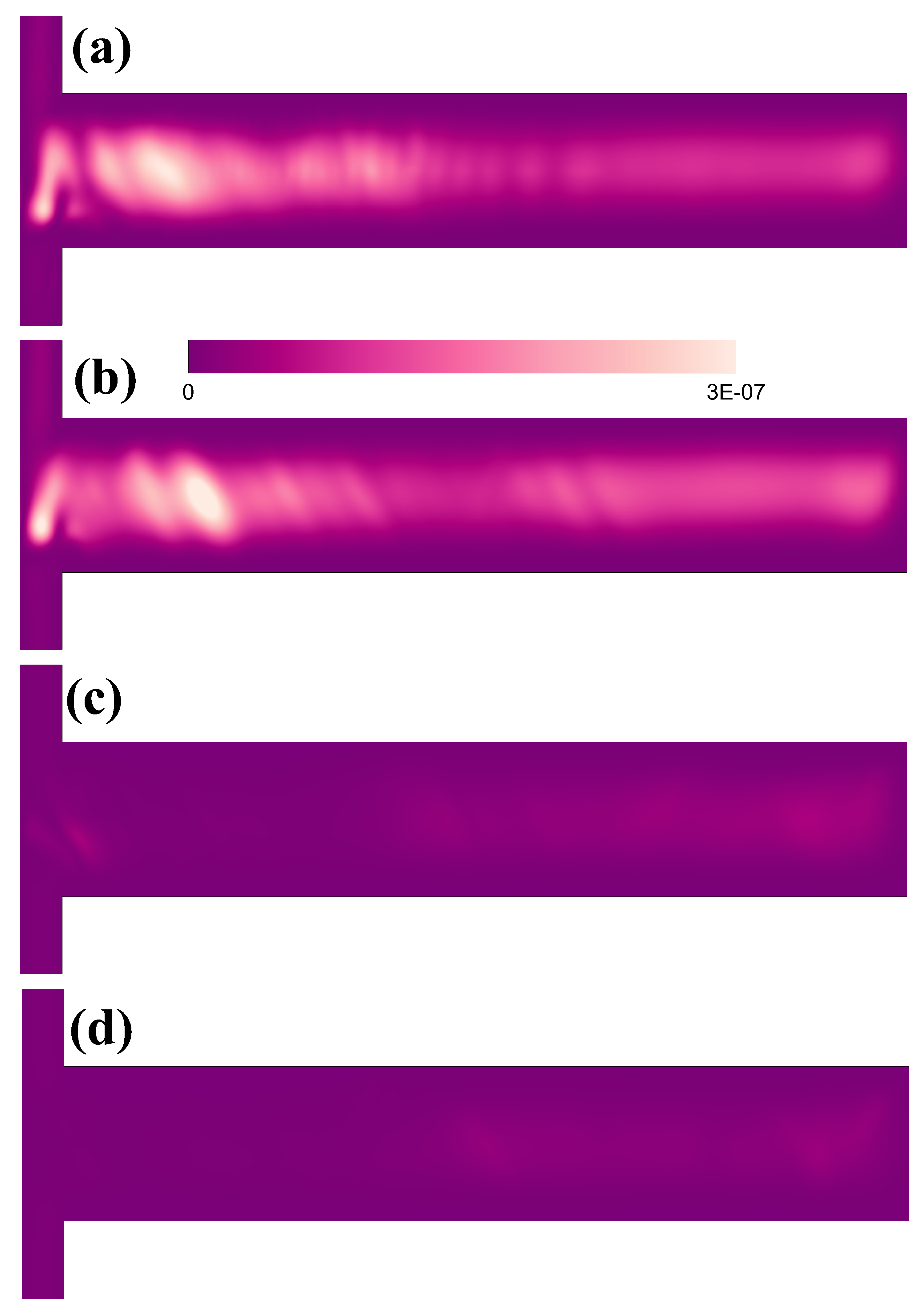}
    \caption{Surface plot of the time-averaged root mean square fluctuation of the span-wise velocity component $u_{2}$ inside the microdevice. (a) Newtonian fluid (b-d) viscoelastic fluids with Weissenberg numbers 1, 10 and 15, respectively.}
    \label{fig5}
\end{figure}

\section{\label{sec:Res}Results and discussion}
The present study performs simulations both for Newtonian and viscoelastic fluids. Simulations were carried out at fixed values of the electrical conductivity ratio of $\gamma = 10$, applied electric field strength of $E_{a} = 20000$ V/m and voltage ratio of $V_{R} = 1$. The values of the electric Rayleigh $(Ra_{e})$ and Reynolds $(Re)$ numbers are kept constant at 2771.6 and 2.77, respectively, as already mentioned in the previous section. To account for the fluid viscoelasticity, we have varied the Weissenberg number between 0 and 15. The case of $Wi = 0$ essentially represents the results of a Newtonian fluid. Note that here the variation in the Weissenberg number is achieved by changing the value of the polymer relaxation time $\lambda$ by keeping everything else at fixed values.       

As mentioned earlier, high and low electrical conductivity fluids enter into the microdevice from the south and north inlets, respectively. They meet at the junction placed at the origin and ultimately leave through the east side of the microdevice. We have presented the instantaneous streamline profiles in Fig.~\ref{fig1} to depict the corresponding velocity profile inside the microdevice at different values of the Weissenberg number. For a Newtonian fluid (sub-Fig.~\ref{fig1}(a)), the streamlines are seen to be highly distorted at the applied condition, thereby showing the existence of a chaotic motion inside the device arising due to the EKI phenomenon at this condition. This distortion of streamlines even leads to the formation of some vortices inside the device. A similar kind of flow behaviour is also observed for viscoelastic fluids with low Weissenberg numbers, for instance, see the results presented at $Wi = 1$ in sub-Fig.~\ref{fig1}(b). This is because, at this low Weissenberg number, the elastic stresses are so low that they can hardly influence the flow behaviour. Therefore, the viscoelastic fluid behaves like a Newtonian one. However, as the Weissenberg number gradually increases to further higher values, a substantial change in the streamline pattern is observed. For instance, at $Wi = 10$, the distortion in the streamline profile is significantly reduced, sub-Fig.~\ref{fig1}(c). The alignment of streamlines is now more ordered than that seen either for Newtonian fluid or viscoelastic fluid with $Wi = 1$. This orderedness in the streamline profile is further accentuated as the Weissenberg number further increases to 15, sub-Fig.~\ref{fig1}(d). At this value of the Weissenberg number, the streamlines are almost straight in the outlet section of the microdevice. This distribution of streamlines at various values of the Weissenberg number signifies that the chaotic motion inside the microdevice diminishes with the increasing value of the Weissenberg number.          

To know about the flow state inside the microdevice, we have plotted the temporal variation of both the non-dimensional stream-wise $(u_{1})$ and span-wise $(u_{2})$ velocities at a probe location placed at $x_{1} = 10H$ and $x_{2} = 0$ in Fig.~\ref{fig2} both for Newtonian and viscoelastic fluids with different Weissenberg numbers. For Newtonian and viscoelastic fluid with $Wi = 1$, both the velocity components exhibit an aperiodic fluctuation, suggesting a chaotic flow state inside the T junction. As the Weissenberg number increases to 10, this fluctuation in the velocity component decreases with time. Also, the nature of the fluctuation transits from an aperiodic to an almost ordered periodic one with low magnitude. On the other hand, at $Wi = 15$, both the velocity components gradually reach a constant steady value with time after some initial fluctuations. Such behaviour in the temporal variation of the probe velocity shows the presence of a steady flow state inside the microdevice at this Weissenberg number. 

To visualize the flow field more clearly, we have presented the instantaneous distribution of the dye concentration inside the microdevice both for Newtonian and viscoelastic fluids in Fig.~\ref{fig3}. Note that here the fluid having high electrical conductivity and entering into the microdevice from the south inlet has a finite non-dimensional dye concentration of $c = 1$, whereas the fluid having low electrical conductivity and entering into the microdevice from the north inlet has a zero concentration, i.e., $c = 0$. A convective-diffusive equation is solved for the evaluation of the dye concentration inside the microdevice along with the governing equations written in the preceding section. The interface of the two fluids (where the gradient of the dye concentration is maximum) becomes highly wavy in nature both for Newtonian and viscoelastic fluid with $Wi = 1$, sub-Figs.~\ref{fig3}(a) and (b), showing the chaotic nature of the flow field inside the microdevice due to the existence of EKI instability. However, this waviness of the dye interface is significantly decreased as the Weissenberg number increases to 10, sub-Fig.~\ref{fig3}(c). At $Wi = 15$, it becomes almost straight in the outlet section of the microdevice.   

All these results suggest that the chaotic or convective motion arising due to the electrokinetic instability is suppressed as the Weissenberg number and/or fluid viscoelasticity gradually increases within the present range of conditions encompassed in this study. A similar observation was also seen in the experimental study of Song et al.~\cite{song2019electrokinetic}. The present study provides a possible explanation for this observation as follows: to explain this, we have plotted the time-averaged and non-dimensional elastic stress component $\tau_{11}$ in Fig.~\ref{fig4} at three different values of the Weissenberg number, namely, 1, 10 and 15. From this figure, one can see the formation of a strand of high $\tau_{11}$ value (known as the birefringent strand), which is started from the origin and then extends up to some distance downstream in the outlet section. It is formed at the interface of the two fluids with high and low electrical conductivities due to the presence of a high extensional flow field in this region. This causes high extension of polymer molecules, which in turn increases the value of the elastic stress component $\tau_{11}$. The presence of such a birefringent strand of high stress value inhibits the onset of electrokinetic instability. This instability originates at the meeting point of the two fluids (slightly away from the origin). It then grows as we move towards the outlet of the device~\cite{chen2005convective}. The chaotic motion arising due to this EKI is mostly caused due to the span-wise fluctuation of the fluid interface. The region of high elastic stress value acts as a barrier to this span-wise flow fluctuation and suppresses the onset of this instability. This is, indeed, evident in Fig.~\ref{fig5} wherein the time-averaged root mean square span-wise velocity fluctuation $u_{rms,2} \,\,(= \sqrt{<\tilde{u_{2}} - \bar{u}_{2}>_{t}})$ is presented both for Newtonian and viscoelastic fluids with different Weissenberg numbers. Here $\tilde{u}_{2}$ is the instantaneous span-wise velocity and $\bar{u}_{2}$ is its time-averaged value. A large span-wise velocity fluctuation is seen both for Newtonian (sub-Fig.~\ref{fig5(a)}) and viscoelastic fluids with $Wi = 1$ (sub-Fig.~\ref{fig5}(b)), which is gradually diminished as the Weissenberg number gradually increases as can be evident from the results presented at Wi = 10 and 15 in sub-Figs.~\ref{fig5}(c) and (d), respectively. At low Weissenberg numbers, for instance, at $Wi = 1$ (sub-Fig.~\ref{fig4}(a)), the formation of such a birefringent strand of high elastic stress value did not occur, and hence the chaotic fluctuation was seen at such Weissenberg number. As the Weissenber number gradually increases, the strength of this strand progressively increases, and hence the resistance in the suppression of the span-wise flow fluctuation also increases. It takes some time to develop this strand of high elastic stress value. As a result, some initial fluctuations were seen even for viscoelastic fluids with high Weissenberg numbers. It was evident from the temporal variation of the probe velocity components presented in Fig.~\ref{fig2}.   

\begin{figure*}[!htbp]
    \centering
    \includegraphics[trim=0cm 0cm 0cm 0cm,clip,width=18cm]{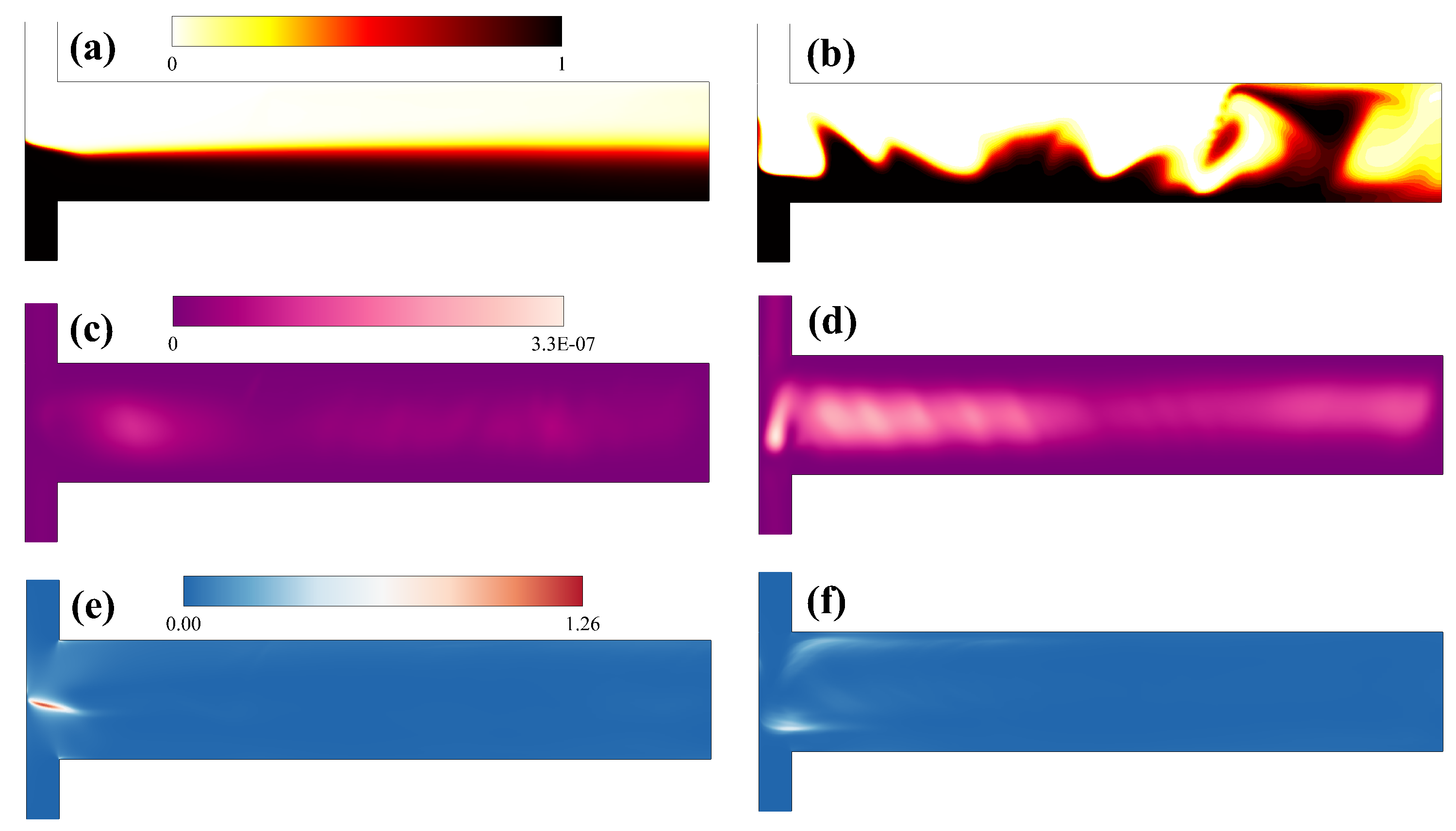}
    \caption{Effect of the polymer viscosity ratio on the flow dynamics. (a) and (b) Instantaneous dye evaluation pattern (c) and (d) Time-averaged root mean square fluctuation of the span-wise velocity component (e) and (f) Variation of the time-averaged elastic stress component $\tau_{11}$. Here the first and second columns represent the results for the polymer viscosity ratios 0.2 and 0.9, respectively.}
    \label{fig6}
\end{figure*}
\begin{figure}[!htbp]
    \centering
    \includegraphics[trim=0cm 0cm 0cm 0cm,clip,width=8cm]{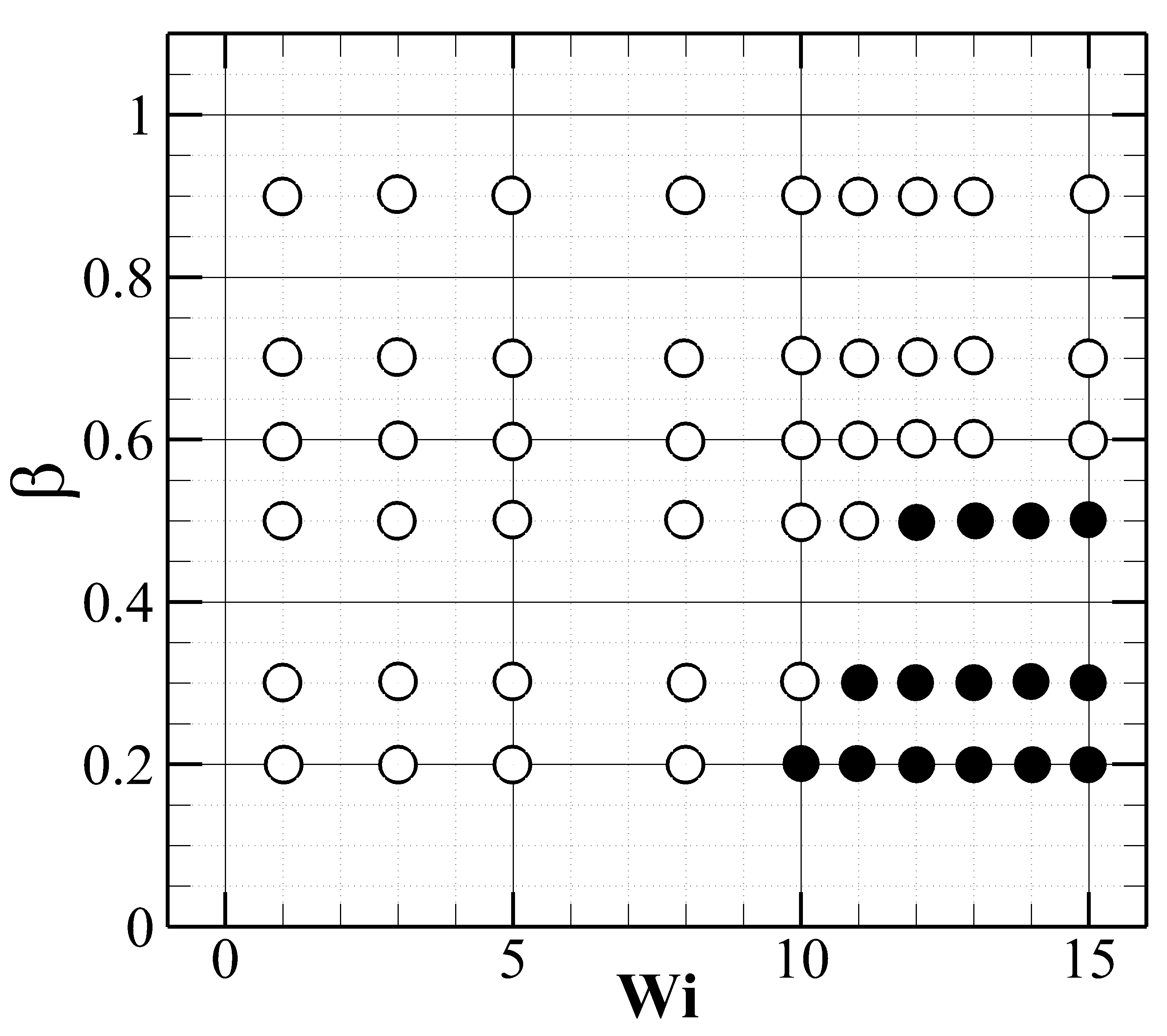}
    \caption{A phase diagram demarcating the presence of steady and unsteady flow regimes on a $\beta-Wi$ space. Here filled and open symbols represent steady and unsteady flow regimes, respectively.}
    \label{phase}
\end{figure}

We have further carried out simulations at various values of the polymer viscosity ratio $\beta$ to examine the influence of the polymer concentration on this EKI instability. Figure~\ref{fig6} depicts the results at two values of the polymer viscosity ratio, namely, 0.2 (first column) and 0.9 (second column) at a fixed value of the Weissenberg number of 15. It can be seen that at this Weissenberg number, the chaotic motion arising due to the EKI instability is suppressed for $beta = 0.2$; likewise, it was observed for $\beta = 0.5$. It is evident from the instantaneous dye evaluation pattern presented in sub-Fig.~\ref{fig6}(a)) wherein the dye interface remains almost straight. However, it becomes highly wavy in nature for $\beta = 0.9$ (sub-Fig.~\ref{fig6}(b)), thereby showing the presence of a chaotic flow field inside the microdevice. It is further apparent in sub-Figs.~\ref{fig6}(c) and (d), wherein the time-averaged root mean square fluctuation of the span-wise $u_{2}$ velocity component is presented for the viscosity ratios 0.2 and 0.9, respectively. The velocity fluctuation is seen to be very less for $\beta = 0.2$ as compared to that seen for $\beta = 0.9$. Therefore, it can be seen that as the value of the polymer viscosity ratio increases or, in other words, as the polymer concentration decreases, the chaotic motion arising due to the EKI instability is augmented. This is because as the polymer viscosity ratio increases, the elastic stresses within the fluid decrease. It is noticeable in sub-Figs.~\ref{fig6}(e) and (f) wherein the time-averaged non-dimensional values of the elastic stress component $\tau_{1}$ are plotted for $\beta = 0.2$ and 0.9, respectively. A strand of high elastic stress is formed at the interface of the two fluids having high and low electrical conductivities for $\beta = 0.2$. It inhibits the span-wise velocity fluctuation, as explained earlier, and hence suppresses the overall chaotic motion inside the microdevice. On the other hand, for $\beta = 0.9$, such a strand of high elastic stress value does not form, and hence the chaotic motion is not suppressed in this case. This dependence of the chaotic motion on the polymer viscosity ratio further demonstrates that the fluid viscoelasticity suppresses it within the range of conditions encompassed in this study. A phase diagram demarcating the steady and unsteady flow regimes is shown in Fig.~\ref{phase} on a $\beta-Wi$ space. It can be seen that as the polymer viscosity ratio decreases and/or the polymer concentration increases, the suppression of the chaotic motion due to the EKI instability happens at earlier values of the Weissenberg number. 

\begin{figure*}
    \centering
    \includegraphics[trim=0cm 0cm 0cm 0cm,clip,width=18cm]{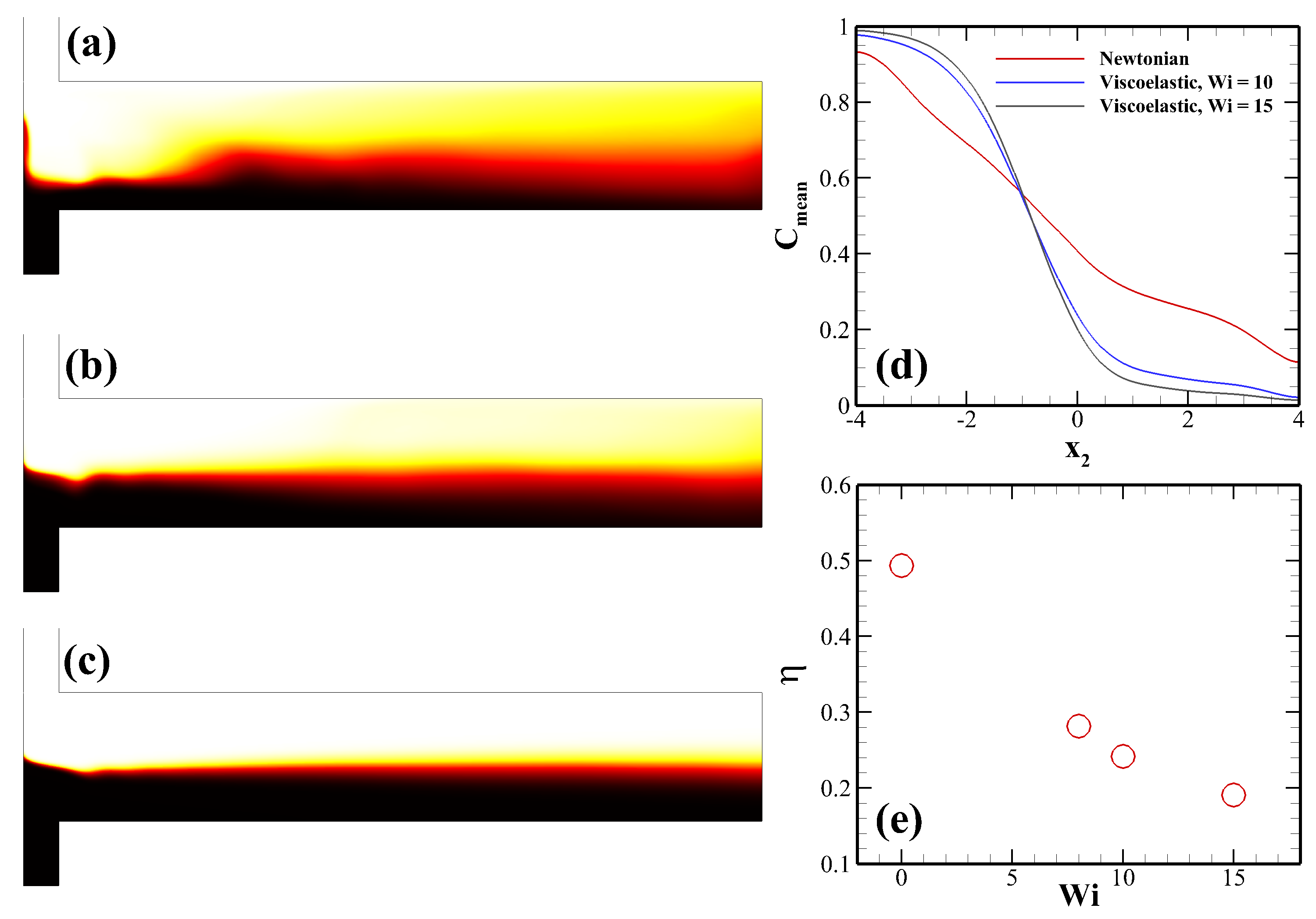}
    \caption{Surface plot of the time-averaged mean dye concentration $(C_{mean})$ profile for (a) Newtonian and viscoelastic fluids with Weissenberg number (b) 10 and (c) 15. (d) Spatial variation of $C_{mean}$ at the outlet of the microdevice both for Newtonian and viscoelastic fluids. (e) Variation of the mixing efficiency index $\eta$ with the Weissenberg number.}
    \label{fig7}
\end{figure*}
Finally, we show how mixing two fluids can be influenced if they are viscoelastic in nature using this EKI phenomenon. For Newtonian fluids, it is well known that the EKI phenomenon always facilitates the mixing~\cite{oddy2001electrokinetic,park2005application,huang2006application,shin2004mixing}. Therefore, it is one of the preferred methods for mixing such fluids in microfluidic devices of various shapes (like T junction, cross-slot, etc.) if they have a gradient in electrical conductivities. However, here we show how this mixing phenomenon can be hampered if the fluids are viscoelastic in nature using this EKI phenomenon. To do so, we have first presented the surface plot of the time-averaged mean dye concentration $C_{mean}$ inside the microdevice both for Newtonian and viscoelastic fluids in Fig.~\ref{fig7}. For Newtonian fluids, one can see that the dye present in the lower half of the fluid spreads to the fluid present in the upper half of the microdevice, particularly at the outlet section, sub-Fig.~\ref{fig7}(a). It occurs due to the mixing of two fluids. However, this spreading of dye reduces for a viscoelastic fluid under the same conditions, and this tendency is accentuated as the Weissenberg number gradually increases; for instance, see sub-Figs.~\ref{fig7}(b) and (c) for the results at Weissenberg numbers 10 and 15, respectively. This is further confirmed in sub-Fig.~\ref{fig7}(d) wherein the variation of $C_{mean}$ is plotted at the outlet of the microdevice along $x_{2}$ direction. For viscoelastic fluids, a steep gradient in the mean dye concentration can be seen around the interface position, suggesting a lower spreading of it from the lower half to the upper half due to lower mixing compared to that for a Newtonian fluid. We have calculated the mixing efficiency $\eta$ to present this mixing phenomenon in a more quantitative manner, which is defined as
\begin{equation}
    \eta = 1 - \frac{\sqrt{\frac{1}{N}\sum_{1}^{N}(C_{mean}^{N} - C_{mean}^{*})^{2}}}{\sqrt{\frac{1}{N}\sum_{1}^{N}(C_{mean}^{0} - C_{mean}^{*})^{2}}}
\end{equation}
where $C_{mean}^{N}$, $C_{mean}^{0}$ and $C_{mean}^{*}$ are the dye concentration at a point along $x_{2}$ direction at the outlet, dye concentration of unmixed fluids and dye concentration of perfectly mixed fluids, respectively. The value of $C_{mean}^{0}$ would be either 0 or 1, and therefore, the value of $C_{mean}^{*}$ would be 0.5. The variation of this mixing efficiency parameter with the Weissenberg number is presented in sub-Fig.~\ref{fig7}(e). It can be seen that the mixing efficiency parameter $\eta$ decreases as we increase the Weissenberg number and/or the fluid viscoelasticity due to the lowering in the chaotic convection inside the microdevice, as discussed earlier.

\section{\label{sec:con}Conclusions}
This study presents a numerical investigation of the electrokinetic instability (EKI) phenomenon in a microfluidic T junction wherein two viscoelastic fluids of different electrical conductivities are transported side by side under the influence of an electric field. For Newtonian fluids, it has been demonstrated by many prior experimental and numerical studies that this EKI phenomenon creates a chaotic flow field inside the microdevice, which can facilitate the mixing of such fluids. Therefore, this method of inducing the EKI phenomenon is considered as one of the promising methods to mix fluids in various micro total analysis systems ($\mu$TAS). However, the present study shows that this method might not work for mixing fluids that are viscoelastic in nature. We have shown that the fluid viscoelasticity (quantified in terms of the Weissenberg number and polymer viscosity ratio) suppresses the chaotic convection inside the microdevice arising due to this EKI phenomenon, which in turn inhibits the mixing of such fluids. Therefore, one has to be careful when planning to mix such viscoelastic fluids using this EKI phenomenon. We have also provided a possible explanation for this behaviour. It is based on the formation of a strand of high elastic stresses at the interface of two fluids, which acts as a barrier to the onset of this chaotic fluctuation. Our observations are in line with that seen in an experiment carried out by Song et al.~\cite{song2019electrokinetic}, who also found a reduction in the wave speed and frequency arising due to this EKI phenomenon in viscoelastic fluids.

\nocite{*}
\bibliography{aipsamp}

\end{document}